\documentclass[letterpaper,10 pt,conference]{ieeeconf}  %
\IEEEoverridecommandlockouts
\usepackage{cite}
\usepackage{amsmath,amssymb,amsfonts}
\usepackage{graphicx}
\usepackage{textcomp}
\usepackage{siunitx}
\usepackage{xcolor}
\def\BibTeX{{\rm B\kern-.05em{\sc i\kern-.025em b}\kern-.08em
    T\kern-.1667em\lower.7ex\hbox{E}\kern-.125emX}}

\title{Dual Skip Connections Minimize the False Positive Rate of Lung Nodule Detection in CT images\\
	\thanks{*Contributed equally}
}

\author{
	\authorblockN{
		Jiahua Xu*,
		Philipp Ernst*,
		Tung Lung Liu*,
		Andreas N\"urnberger
	}\\
	\authorblockA{
		Data \& Knowledge Engineering Group, Faculty of Computer Science\\
		Otto von Guericke University Magdeburg, Germany\\
		\{Jiahua.Xu, Philipp.Ernst, Andreas.Nuernberger\}@ovgu.de \\ {tung-lung.liu}@st.ovgu.de
	}
}

\begin{document}
	
\begin{titlepage}
\thispagestyle{empty}
\noindent
{\Large Dual Skip Connections Minimize the False Positive Rate of Lung Nodule Detection in CT images}\\\\
{\large Jiahua Xu*, Philipp Ernst*, Tung Lung Liu*, Andreas N\"urnberger}\\
{*contributed equally}\\\\
Data \& Knowledge Engineering Group, Faculty of Computer Science\\
Otto von Guericke University Magdeburg, Germany\\\\
\{Jiahua.Xu, Philipp.Ernst, Andreas.Nuernberger\}@ovgu.de \\ {tung-lung.liu}@st.ovgu.de\\
\vspace{2cm}\\
{\small Pre-print version}
\vspace{\fill}\\
\textcopyright{} 2021 IEEE. Personal use of this material is permitted. Permission from IEEE must be obtained for all other uses, in any current or future media, including reprinting/republishing this material for advertising or promotional purposes, creating new collective works, for resale or redistribution to servers or lists, or reuse of any copyrighted component of this work in other works.
\end{titlepage}
\newpage

\maketitle

\begin{abstract}
Pulmonary cancer is one of the most commonly diagnosed and fatal cancers and is often diagnosed by incidental ﬁndings on computed tomography. Automated pulmonary nodule detection is an essential part of computer-aided diagnosis, which is still facing great challenges and difficulties to quickly and accurately locate the exact nodules' positions. This paper proposes a dual skip connection upsampling strategy based on Dual Path network in a U-Net structure generating multiscale feature maps, which aims to minimize the ratio of false positives and maximize the sensitivity for lesion detection of nodules. The results show that our new upsampling strategy improves the performance by having 85.3\% sensitivity at 4 FROC per image compared to 84.2\% for the regular upsampling strategy or 81.2\% for VGG16-based Faster-R-CNN.
\end{abstract}


\section{Introduction}
Pulmonary cancer is one of the most commonly diagnosed and fatal cancers among other cancers in medical research \cite{ferlay2015cancer}. Other pulmonary diseases, such as COVID-19 or pulmonary infection, may also cause serious damage to the lung. The most common problem during diagnostic in radiology is solitary pulmonary nodules, i.e. single, round or oval nodules generally smaller than \SI{3}{\centi\meter} \cite{Toghiani.2015}.

Computed tomography (CT) is one of the most common non-invasive screening approaches for diagnosing pulmonary diseases \cite{Midthun.2016}. Pulmonary nodules that appear on the images have a high variability in terms of size, shape and location in the pulmonary regions \cite{rubin2015lung}. Small nodules are very difficult to observe because there are many other tissues in the thorax (e.g., blood vessels, airways, lymph nodes) with morphological features similar to nodules \cite{torre2016lung}. It is challenging to reduce misdiagnoses and false positives (FPs) in early-stage lung cancer diagnosis \cite{dou2016multilevel}. The main issue which leads to a high ratio of FPs, particularly for pulmonary nodule detection, comes from the variability of nodules in terms of size, shape and location \cite{rubin2015lung} and, compared to regular RGB images, gray-scale medical images provide less information in terms of edges and textures to distinguish different tissues \cite{escobar2008interactive}. 

In recent researches, deep learning  has greatly improved performance and efficiency in nodule detection. Variations based on U-Net \cite{8648753}, Faster R-CNN \cite{ding2017accurate}, 3D-CNN\cite{zhu2018deeplung}, and 3D-Dual Path Network \cite{jiang2020attentive} have been reported. One study \cite{yan20183d} utilized neighboring slices to extract the volumetric and contextual information around the nodules as well as keeping the computational effort of the method low compared to 3D models that have a larger number of network parameters. 

In this paper, we propose a dual skip connection upsampling strategy using Faster R-CNN \cite{Ren.2015} with Dual Path Network (DPN) \cite{Chen.2017} as the backbone in a U-Net \cite{Ronneberger.2015} structure. 

\section{Method}
\subsection{Dataset: DeepLesion}
DeepLesion \cite{Yan.2018} was released by the National Institutes of Health Clinical Center. It consists of 32,120 axial CT slices from 10,594 CT scans (studies) of 4,427 unique patients. We have extracted CT images that are annotated with pulmonary nodules, which resulted in 2,394 CT images for our dataset in our case. 1,916 CT images are used for training and 478 CT images for validation. 

\subsection{Data Preprocessing}
The CT slices of DeepLesion are normalized by converting Hounsfield units to mass attenuation coefficients and dividing by the 99th percentile of the entire dataset. Each slice has \SI{1}{\milli\meter} to \SI{5}{\milli\meter} thickness in most cases while some of the images are \SI{0.625}{\milli\meter} or \SI{2}{\milli\meter}. For each lesion, there is one key slice with \SI{30}{\milli\meter} of extra slices in front of and behind the key slice. However, only the key slice has the annotation data including lesion types, coordinates of 2D bounding-boxes and RECIST diameters for the lesions. The CT slices were resized to \SI{512x512}{px}. For training, we adapted data augmentation where images are flipped horizontally and vertically and rotated with a probability of 50\% respectively to enrich the variability of the CT images. During augmentation, the corresponding coordinates of the bounding boxes are updated as well. To enhance the spatial information, we concatenate one more slice in front of and behind the key slice to get a 2.5D model. This approach increases the information for the model while being a lot more lightweight than a 3D model.  

\subsection{Proposed Model}
\begin{figure*}
\centering
\includegraphics[width=0.85\textwidth]{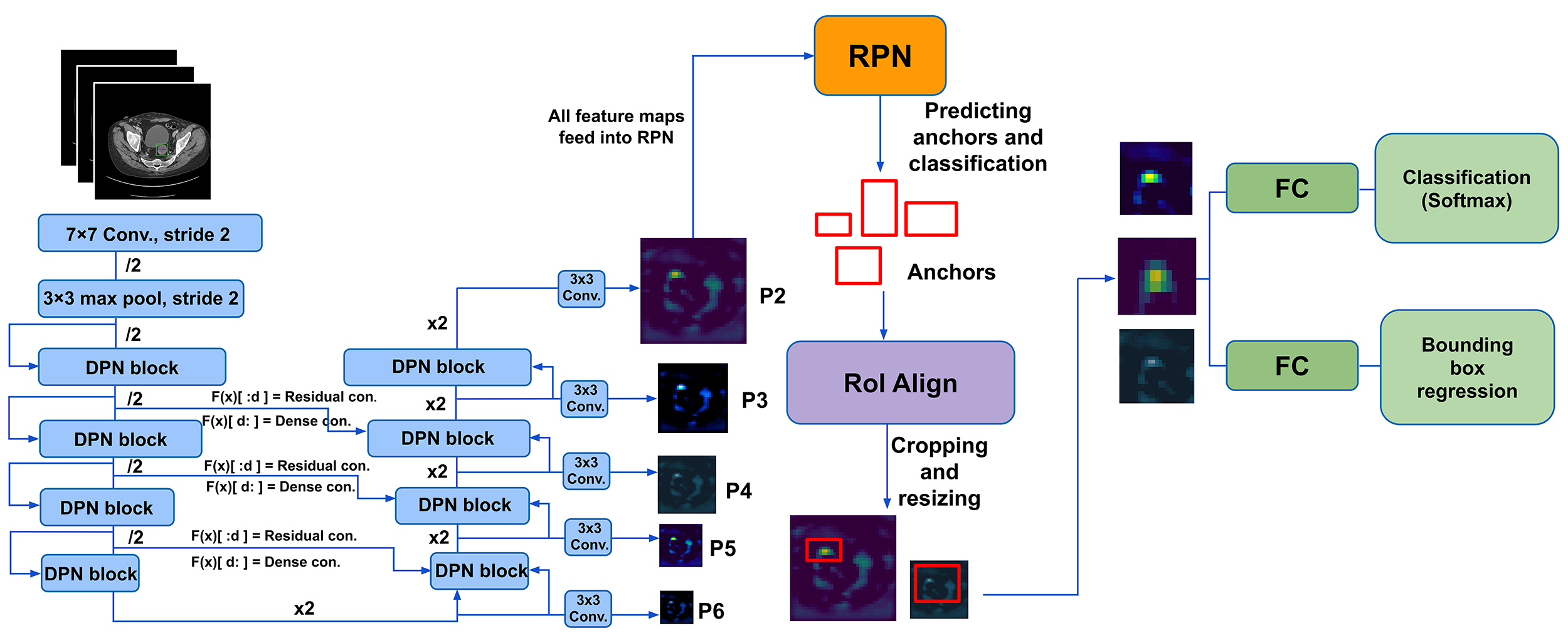}
\caption{Our model generates five feature maps for RPN. ROI Align crops the feature maps according to the sizes of the anchors on the corresponding feature map.}
\label{proposed_model}
\end{figure*}
We adopt ideas from different algorithms to tackle the issue of variability of pulmonary nodules. The two-stage object detection model Faster R-CNN is used, which is hoped to reduce the ratio of FPs by providing more spatial and contextual information. Fig.~\ref{proposed_model} shows the outline of the architecture of the proposed model.

The backbone architecture of the first stage is a combination of U-Net and DPN with skip connections between the encoding and decoding path to provide more spatial information and to improve the flow of gradients.
Five feature maps of different resolutions are derived during the decoding process that are fed into a Region Proposal Network (RPN) \cite{Ren.2015} for the first part of classification and anchor box regression.

ROI Align from Mask R-CNN \cite{he2017mask} crops and resizes the regions of interest of RPN to a fixed resolution. Eventually, two sets of fully connected layers, which classify nodule or non-nodule, are attached at the end of our network for the second part of classification and anchor box regression, respectively. 

\subsubsection{Encoder}
The encoder is a modified DPN architecture that performs DenseNet and ResNet in parallel. The model starts with having 64 kernels with a 7x7 convolutional layer and a stride of 2 followed by a 3x3 max pooling layer with a stride of 2 subsequently as an initial block. Afterwards, there are four more bottleneck blocks \cite{Chen.2017} having 1x1, 3x3 and 1x1 convolutional layers where each convolutional layer is followed by batch normalization and ReLU activation. Grouped convolutions on all channels are performed within the 3x3 convolutional layer like in ResNet \cite{xie2017aggregated}.

\subsubsection{Decoder}
\begin{figure}
\centering
\includegraphics[width=\linewidth]{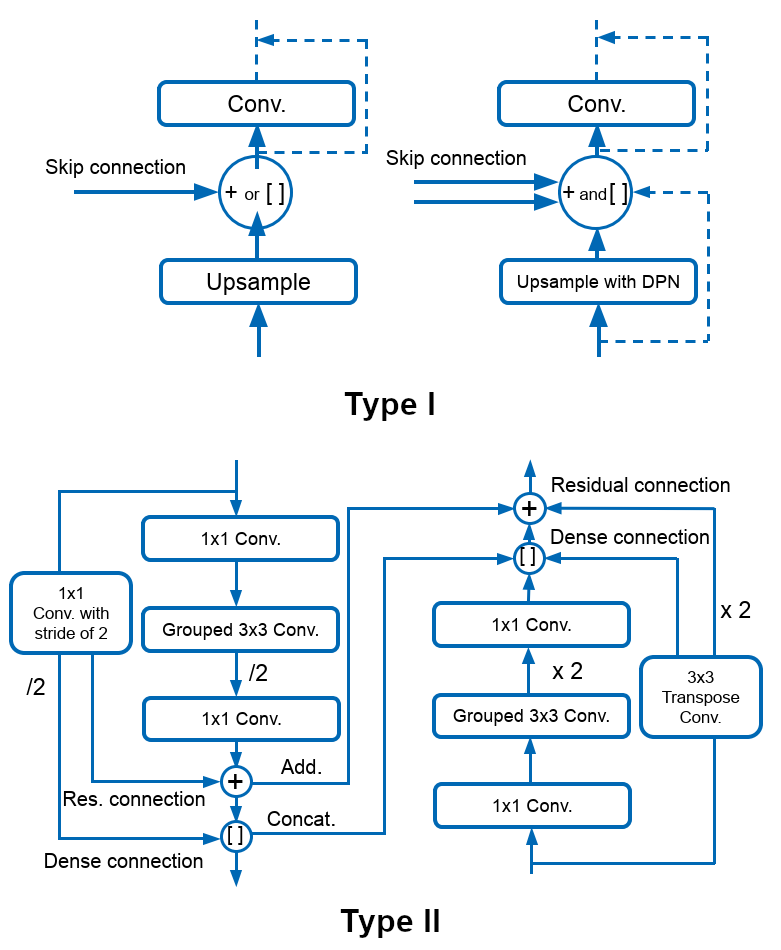}
\caption{ Upsampling Type \MakeUppercase{\romannumeral1} and Type \MakeUppercase{\romannumeral2}}
\label{sampling}
\end{figure}

The decoder stages consist of scaling up the feature maps and concatenating or adding the skip connections in the same encoder stage, where the upsampling starts after scaling up the feature map, as shown in Fig.~\ref{sampling} (Type \MakeUppercase{\romannumeral1}). This approach is relatively straightforward and easy to implement since ResNet-like, DenseNet-like or FCNs have only a single type of operation to combine the different connections, yet, DPN has both of these operations during encoding.
The implementation of starting the shortcut before the upsampling layer is shown in Fig.~\ref{sampling} (Type \MakeUppercase{\romannumeral2}). Hence, it provides us a discussion space to observe the performance if we start the shortcut before upsampling and implement both addition and concatenation to connect with the skip connection as a regular DPN block during decoding. Finally, an extra 3x3 convolutional layer is attached after concatenating or adding the skip connections.

We have implemented two different types of upsampling with DPN blocks as the backbone to observe the performance on pulmonary nodule detection. Type \MakeUppercase{\romannumeral1} is a regular upsampling approach and starts the DPN block after upsampling as shown in Fig.~\ref{sampling}. For Type \MakeUppercase{\romannumeral2}, we extract a part of the data stream before upsampling, where the main data stream is upscaled in resolution by a DPN block. This is done by the 3x3 convolutional layer within the bottleneck in a DPN block with a stride of 2. The detailed visualization of our proposed upsampling strategy is depicted in Fig.~\ref{sampling}, which shows how the two operations of concatenation and addition interact during encoding and decoding.

\subsection{Generating Ground Truth Labels}
Considering each scale of an anchor with three different ratios is performed on one single level of the feature map, the concatenation for training labels is not applicable because the height and width are different on each level. For this reason, we vectorize all labels and concatenate them. However, this approach requires a lot more attention to making sure the order of the labels is valid in the order of how convolutional filters are doing windowing on feature maps from different levels.

For RPN, we mark an anchor as positive if the IoU is more than 0.7 with the ground truth bounding box and as negative if the IoU is lower than 0.3. Those anchors having IoU between these maximum and minimum threshold are considered as neutral, and are not involved for training.

The sub-network Classifier relies on the prediction of RPN to provide information on the location of the target. We consider locations as background if the IoU is between 0.3 and 0.5 and foreground if the IoU is greater than 0.5. Only positive samples calculate the regression parameters in this stage as well.

\subsection{Loss Function}
Our loss function follows the definition of the multi-task loss function of Faster R-CNN (cf. Eq.~(1-2) in \cite{Ren.2015}), i.e.
\begin{equation*} \label{Loss}
L(p,t) = \frac{1}{N_{cls}} \sum L_{cls}(p,p^{\ast})+\lambda
\frac{1}{N_{reg}} \sum p^{\ast}L_{reg}(t,t^{\ast}),
\end{equation*}
consisting of: 
\begin{itemize}
\item The classiﬁcation loss $L_{cls}$, the $\log$-loss over two classes (object vs. not object) for RPN and multiple classes for Classifier.
\item The regression loss $L_{reg}$, that we set to the smooth L1 loss \cite{Girshick.2015}.
\end{itemize}
The term $p^{\ast}L_{reg}$ calculates the regression loss only for positive anchors $(p^{\ast}=1)$. The weighting parameter $\lambda$ is set to 1, since it was proven insensitive in \cite{Ren.2015}. For bounding box regression, we adopt the regression parameters of the 4 coordinates following R-CNN's definition (cf. Eq.~(6-9) in \cite{girshick2014rich}).

\subsection{Metric}
The free-response receiver operating characteristic (FROC) is one of the standard metrics in lesion detection \cite{setio2017validation}. Its evaluation is performed by measuring the sensitivities (\%) with respect to their corresponding average FP rate per scan. TPs and FPs are determined by thresholding a confidence measure of the predictions. For our evaluations, we calculate the IoU of the predicted bounding boxes with the ground truth bounding boxes. If it is larger than 0.5, it represents a TP, or an FP otherwise.

\subsection{Experimental Setup}
For training, we use Adam as optimizer with its default parameters (learning rate=\SI{1e-3}{}, betas=(0.9, 0.999), eps=\SI{1e-08}{}) and a weight decay of \SI{1e-4}{}. The initial learning rate is reduced to \SI{1e-4}{} after 5 epochs. Due to time limitations, all models are trained within 15 epochs on the training dataset. We use data augmentation, dropout and normalization to prevent overfitting. The initialization of the layers’ parameters is also kept at default, i.e. Xavier uniform for kernels and zeros for biases. All trainings and tests are performed on Google Colab Pro (NVIDIA Tesla K80) utilizing Keras 2.3.1 with tensorflow as backend. 

\section{Results}
We implemented three different models. Our baseline model is based on Faster R-CNN with VGG16 as backbone where the prediction was only performed on the last feature map. The other two models are our proposed models with the regular upsampling strategy (Type \MakeUppercase{\romannumeral1}) and the proposed upsampling approach (Type \MakeUppercase{\romannumeral2}).

Tab.~\ref{FROC} shows the the sensitivities at 1/2, 1, 2, 4, 8 and 16 average FPs per scan to compare our models. At 4 average FPs per scan, e.g., the sensitivity of our model using Type \MakeUppercase{\romannumeral2} upsampling is increased by \SI{1.1}{\percent} compared to Type \MakeUppercase{\romannumeral1} upsampling, and is even \SI{4}{\percent} above the Faster R-CNN baseline.

\begin{figure}
\centering
\includegraphics[width=0.42\textwidth]{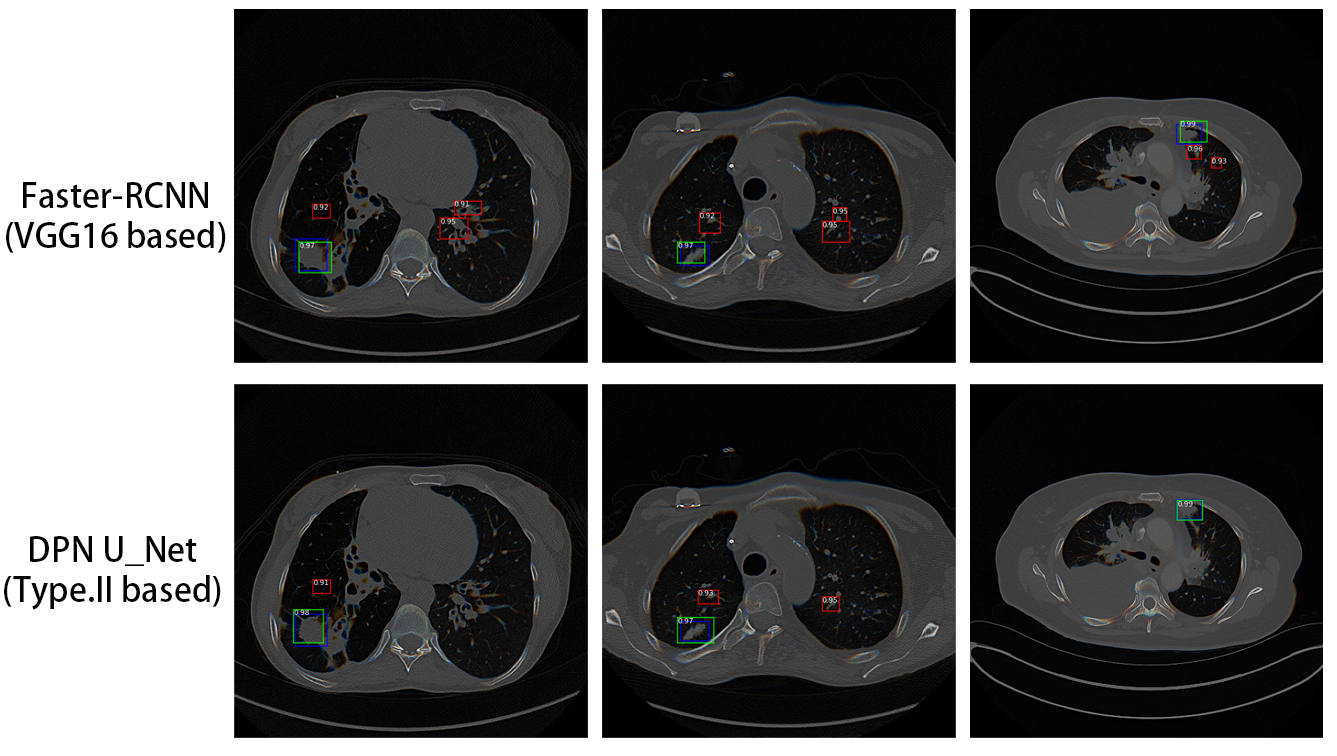}
\caption{Exemplary predictions on the CT slices using VGG16 based Faster-RCNN and DPN U-Net (Type \MakeUppercase{\romannumeral2}).}
\label{VIS_DPUet}
\end{figure}

The first row of Fig.~\ref{VIS_DPUet} visualizes the detection results in the official test dataset of DeepLesion for the baseline model while the second row of Fig.~\ref{VIS_DPUet} shows the results of the proposed model DPN U-Net Type \MakeUppercase{\romannumeral2}. Blue, green and red boxes represent the ground truth, TP and FP boxes respectively, and the number on the top-left corner of the boxes represents the confidence.

\section{Discussion}
From Tab.~\ref{FROC}, we see that the average FROC of DPN U-Net Type \MakeUppercase{\romannumeral2} yields 80.5\%, surpassing the DPN U-Net Type \MakeUppercase{\romannumeral1} with 78.4\%, and the baseline model with 75.1\%. The baseline model is a single scale model where the RPN only looks for targets on the same resolution of the feature map. By looking at Fig.~\ref{VIS_DPUet}, we can see that the baseline model generates more FP predictions than DPN U-Net Type \MakeUppercase{\romannumeral2} on the same CT images in general. Our results show that multiscale feature maps can help to improve the performance.

\begin{table}
\footnotesize
\caption{Sensitivities at different FPs per image (multiple lesions).}
\centering
\begin{tabular}{lcccccc}
\hline
 & \multicolumn{6}{c}{Sensitivity (\%) at FPs}\\
Model & 0.5 & 1 & 2 & 4 & 8 & 16\\ \hline 
\textbf{Faster R-CNN (VGG16)} & 55.8 & 66.3 & 74.7 & 81.2 & 84.8 & 87.5\\
\textbf{DPN U-Net, Type \MakeUppercase{\romannumeral1} } & 60.9 & 71.3 & 77.7 & 84.2 & 87.6 & 88.9\\
\textbf{DPN U-Net, Type \MakeUppercase{\romannumeral2} } & 64.6 & 74.1 & 80.7 & \textbf{85.3} & 88.3 & 89.8 \\
\hline
\end{tabular}
\label{FROC}
\end{table}
The traditional upsampling approach in Type \MakeUppercase{\romannumeral1} might lose some information when upsampling, although it has the skip-connection in the same stage. DPN U-Net Type \MakeUppercase{\romannumeral2} intends further to provide more contextual information upon the original structure. The results show that Type \MakeUppercase{\romannumeral2} is an efficient approach to reuse the DPN block and can provide more contextual information by adding a shortcut connection before upsampling. Furthermore, during the experiment, DPN U-Net Type \MakeUppercase{\romannumeral1} and \MakeUppercase{\romannumeral2} require a similar computational time per batch, while the baseline model requires only 50\% of the time for computation. This behavior is expected since DPN consists of more complex operations and structure.

We also compared our proposed models with the state of art, such as a 3DCE\_CS\_Att network \cite{tao2019improving} with 21 slices for pulmonary nodule detection achieving a sensitivity of 92\% at 4 FPs while having average FROC of 83.9\% for all lesions from DeepLesion. Yet, the sensitivity of DPN U-Net Type \MakeUppercase{\romannumeral2} at 4 FPs is at 85.3\%, which is already quite close to 3DCE with 89\%. By theory, 3D input provides more contextual information for a deep learning model, yet, it also requires more computational resources.

\section{Conclusion}
This paper proposed a dual skip connection upsampling strategy to locate pulmonary nodules in various shapes and sizes compared with two baseline networks. Our work shows that the proposed model DPN U-Net Type \MakeUppercase{\romannumeral2} surpasses the results performed by the single skip connections model (Type \MakeUppercase{\romannumeral1}) and single-scale feature map model (Faster R-CNN). The proposed model DPN U-Net Type \MakeUppercase{\romannumeral2} reuses the DPN block throughout the whole network, which is an efficient way to explore new potential features and prevent vanishing gradients by having both operations from ResNet and DenseNet. Overall, our proposed upsampling strategy has successfully reduced the false positives in the evaluation of nodule detection. We assume that the performance of our proposed model might still be improved by fine tuning the hyperparameters. 

\section*{Acknowledgement}
This work was conducted within the International Graduate School MEMoRIAL at OVGU Magdeburg, supported by the ESF (project no. ZS/2016/08/80646).

\bibliography{conference_041818.bib}
\bibliographystyle{IEEEtran}

\end{document}